\documentclass{article}
\usepackage{graphicx} 
\usepackage{amsmath}
\usepackage{amsthm}
\usepackage{amssymb}
\usepackage{mathtools}
\usepackage{xcolor}
\usepackage[round]{natbib}
\usepackage{enumerate}
\usepackage{bbm}
\usepackage{bm}
\usepackage{physics}
\usepackage{apptools}
\usepackage{float}
\usepackage{threeparttable}
\usepackage{rotating} 
\usepackage{enumitem}
\usepackage{subfig}
\usepackage{placeins}

\usepackage[a4paper, total={6in, 8in}]{geometry}

\usepackage{authblk}

\usepackage{orcidlink}

\AtAppendix{\counterwithin{theorem}{section}}

\newtheorem{theorem}{Theorem}
\newtheorem{lemma}[theorem]{Lemma}
\newtheorem{remark}{Remark}
\newtheorem{corollary}[theorem]{Corollary}

\newtheorem{algorithm}{Algorithm}

\theoremstyle{definition}

\providecommand{\keywords}[1]
{
  \small	
  \textbf{\textit{Keywords---}} #1
}

\newcommand\extrafootertext[1]{%
    \bgroup
    \renewcommand\thefootnote{\fnsymbol{footnote}}%
    \renewcommand\thempfootnote{\fnsymbol{mpfootnote}}%
    \footnotetext[0]{#1}%
    \egroup
}
\newcommand{\lebesgue}{\ensuremath{\lambda\!\!\lambda}}

\title{Inference for competing risks based on area between curves statistics}

\author{
Simon Mack $^{1,\ast}$\orcidlink{0009-0006-0081-935X},
Marc Ditzhaus $^{2, \dagger}$\orcidlink{0000-0001-9235-1905},
Merle Munko $^{2}$\orcidlink{0009-0002-6142-8142} and
Markus Pauly $^{3,4}$\orcidlink{0000-0002-0976-7190}}

\affil{
$^{1}$ Institute of Statistics, RWTH Aachen University, Aachen, Germany \\
$^{2}$ Faculty of Mathematics, Otto von Guericke University Magdeburg, Magdeburg, Germany\\
$^{3}$ Department of Statistics, TU Dortmund University, Dortmund, Germany\\
$^{4}$ Research Center Trustworthy Data Science and Security, UA Ruhr, Germany}

\date{\today}

\begin{document}

\maketitle

\begin{abstract}
In competing risks models, cumulative incidence functions 
are commonly compared to infer differences between groups. Many existing inference methods, however, struggle when these functions cross during the time frame of interest. To address this problem, we investigate a test statistic based on the area between cumulative incidence functions. As the corresponding limiting distribution depends on  quantities that are typically unknown, we propose a wild bootstrap approach to obtain a feasible and asymptotically valid two-sample test. The finite sample performance of the proposed method, in comparison with existing methods, is examined in an extensive simulation study.
\end{abstract}


\extrafootertext{$\ast$ Corresponding author. Email adress: \texttt{simon.mack@rwth-aachen.de}}
\extrafootertext{$\dagger$ Deceased on September 11, 2024.}
\section{Introduction}
The analysis of time-to-event data plays an important role in various fields such as medicine or engineering, e.g. when different treatment methods are compared in clinical trials, or the lifetime of industrial components is estimated.

In applications, two-sample comparisons are often of primary interest and in the simple survival setting with independent right-censored event times many hypothesis tests for euqality are available. Many of these tests are based on the generalized log-rank statistic \citep{gehan1965generalized,mantel1966evaluation,peto1972asymptotically} and therefore utilize the one-to-one relationship between survival functions and hazard rates. 

While classical survival analysis typically focuses on a single event type, many applications naturally lead to competing risks settings. In such situations, several possible causes of failure can occur. In oncology, e.g., death due to cancer is usually the event of interest, whereas other causes of death like surgical mortality are competing events \citep{putter2007tutorial}. In competing risks settings, the cumulative incidence functions (CIFs) are generally of primary interest. Because each CIF usually depends on all cause-specific hazard functions, generalizing the construction of the log-rank test to competing risks data is not possible directly. 

The first test for the equality of CIFs between two (or more) samples was proposed by \citet{gray1988class} and is based on subdistribution hazards, which lack the intuitive interpretation of cause-specific hazard rates. The first two-sample test based directly on the CIFs was proposed by \citet{pepe1991inference}. 
Both tests, however, struggle to detect differences between groups if the CIFs of the cause of interest cross, due to the construction of the test statistics. Such situations may occur in cancer trials, e.g. when comparing chemotherapy with monoclonal antibodies \citep{ferris2016nivolumab}. 

As alternatives, distance-based tests of Kolmogorov-Smirnov \citep{lin1997non} and Cramér-von-Mises \citep{dobler2017non} type are available. Even though these methods yield omnibus tests, they are not specifically designed to handle crossing CIFs. To address this problem, \citet{lyu2020comparison} adopted an approach of \citet{lin2010new} and proposed a test statistic, which is based on the area between the CIF curves. \citet{lyu2020comparison} postulated asymptotic normality of their test statistic but they did not provide a proof of this statement. 

In this paper, we focus on the two-sample comparison of CIFs, with particular attention to crossing curves. First, we identify a flaw in the claimed asymptotic normality of the area based test statistic of \citet{lyu2020comparison}. As we prove, the asymptotic distribution is more complex and depends on quantities that are typically unknown. A similar issue regarding the test of \citet{lin2010new} was already pointed out and corrected by \citet{liu2020resampling} in the simple survival case. As a feasible alternative, we propose a wild bootstrap test, and prove that it shares the same desirable asymptotic properties as the original test.

The remainder of the paper is organized as follows. Section 2 introduces the competing risks model and the testing problem of interest. In Sections 3 and 4, we present our proposed test and its bootstrap version. Section 5 contains an extensive simulation study comparing our test with some existing methods, indicating comparable performance for ordered alternatives and superior power  in the presence of crossing CIFs. Section 6 illustrates the application of our test to bone marrow transplantation data, before summarizing our findings in Section 7. All proofs are given in the Appendix, which also contains details regarding a possible discontinuity adjustment as well as additional simulation results.

\section{Model and Notation}
We consider a two-sample competing risks setup, 
where each group $j, (j=1,2)$ is modeled by individual competing risk processes  $(X^{(j)}(t))_{t\ge0}$ for   with $m\in \mathbb{N}$ competing risks. We assume those processes to be càdlàg. They are non-homogeneous Markov chains with state space $\{0,1, \dots, m\}$ and initial state $0$, i.e., $P(X^{(j)}(0) = 0) = 1$. The other states $1, \dots, m$ represent our competing events and are absorbing.  For ease of presentation, we limit ourselves to the consideration of only $m=2$ risks. This is not a practical limitation, as in most applications only one cause (here, the first) is of interest, while all other causes of failure can be combined into a single competing event (the second), which is then treated as a nuisance parameter.
The event time is the transition time out of the initial state $T^{(j)}=\inf\{t>0 | X^{(j)}(t) \ne 0\}$, and we denote by $X^{(j)}_T \in \{1,2\}$ the event type. The transition rates out of the initial state are given by the cause-specific hazard functions
\begin{equation*} \label{hazards}
\alpha^{(j)}_k(t) = \lim_{\varepsilon \to 0^+} \frac{P(T^{(j)} \in [t, t+ \varepsilon ], X^{(j)}_T = k | T \ge t)}{\varepsilon}, \quad k=1,2
\end{equation*}
which we assume to exist. They completely determine the distribution of  $T^{(j)}$ due to the relation
\begin{equation*} \label{distT}
	 P(T^{(j)} > u) = \exp(-\int_{0}^{u}(\alpha^{(j)}_1(s) + \alpha^{(j)}_2(s))ds), \quad u \in [0, \infty].
\end{equation*}
Throughout we assume that the observation of $(T^{(j)}, X^{(j)}_T)$ is subject to independent left-truncation and right-censoring. This includes, but is not limited to, random left-truncation and right-censoring, where $(T^{(j)}, X^{(j)}_T)$ is independent of a random pair of truncation and censoring times. For details regarding the modeling of incomplete observations we refer to Chapter III of \citet{andersen1993statistical}. Denote by $n_j^{\prime} \in \mathbb{N}, j=1,2$  the number of underlying individuals in each group and let $L^{(j)}_i$ and $C^{(j)}_i$, $i=1,\dots,n_j^{\prime}$ denote the individual left-truncation and right-censoring times, respectively. We define the observed time by $X^{(j)}_i=\min(T^{(j)}_i, C^{(j)}_i)$ and the event indicator by $\delta^{(j)}_i= X^{(j)}_{T_i}$ if $X^{(j)}_i=T^{(j)}_i$ and $\delta^{(j)}_i=0$ if $X^{(j)}_i=C^{(j)}_i$. We therefore only observe the $n_j \le n_j^{\prime}$ triplets $(L^{(j)}_i,X^{(j)}_i,\delta^{(j)}_i)$ satisfying $L^{(j)}_i < X^{(j)}_i$.

Our main quantities of interest are the cumulative incidence functions (CIFs), also called subdistribution functions
\begin{equation*}
	F^{(j)}_k(t) = P(T^{(j)} \le t, X^{(j)}_T=k) = \int_{0}^{t}P(T^{(j)} > u-)\alpha^{(j)}_k(u)du,\quad k=1,2.
\end{equation*}
 Due to \eqref{distT} the CIFs are deterministic functions of the cause-specific hazards. 
 Following the counting process framework of \citet{andersen1993statistical}, we define
\begin{equation*}
	Y^{(j)}_i(t)= \mathbbm{1}(\text{individual $i$ in group $j$ is observed to be in state $0$ just before $t$}),
\end{equation*}
\begin{equation*}
	N^{(j)}_{k,i}(t) = \mathbbm{1}(\text{individual $i$ in group $j$ has observed $0\to k$ transition in $[0,t]$}),
\end{equation*}
where $i=1, \dots, n_j, k=1,2, j=1,2$ and $\mathbbm{1}(\cdot)$ is the indicator function. Furthermore let for each group $Y^{(j)}(t)=\sum_{i=1}^{n_j}Y^{(j)}_i(t)$ denote the number of individuals at risk just before $t$, $N^{(j)}_{k}(t)=\sum_{i=1}^{n_j}N^{(j)}_{k,i}(t)$ the observed events of type $k$, and $N^{(j)}(t)=N^{(j)}_{1}(t)+N^{(j)}_{2}(t)$ all observed events in the interval $[0,t]$.
 Since independent left-truncation and right-censoring can be modeled by an {Aalen filter}, see \citet[§III.4]{andersen1993statistical}, the counting processes $N^{(j)}_{k,i}(t)$ follow the multiplicative intensity model with intensity 
\begin{equation*}
	\lambda^{(j)}_{k,i}(t)=Y^{(j)}_{i}(t)\alpha^{(j)}_k(t).
\end{equation*}
The corresponding compensated processes 
\begin{equation*} \label{martingale}
	M^{(j)}_{k,i}(t)= N^{(j)}_{k,i}(s) - \int_{0}^{t}Y^{(j)}_{i}(s)\alpha^{(j)}_k(s)ds
\end{equation*}
are therefore local square integrable martingales, see Proposition II.4.1 of \citet{andersen1993statistical}.
 
The unknown CIFs can be estimated by the Aalen-Johansen estimators \citep{aalen1978empirical}
 \begin{equation*} \label{AJest}
 	\hat{F}^{(j)}_k(t) = \int_{0}^{t} \frac{\hat{K}^{(j)}(u-)}{Y^{j}(u)} dN^{(j)}_{k}(u),\quad t \ge 0,  
 \end{equation*}
where $\hat{K}^{(j)}(u) = \prod_{s\le u}(1-\frac{\varDelta N^{(j)}(s)}{Y^{(j)}(s)})$ is the Kaplan-Meier estimator \citep{kaplan1958nonparametric} for the overall survival function $K^{(j)}(u)=P(T^{(j)} > u)$. Here $\varDelta N^{(j)}(s)=N^{(j)}(s)-N^{(j)}(s-)$ counts the number of all events at time $s$ in group $j$.  Furthermore the integrand in \eqref{AJest} is interpreted to be $0$ if $Y^{(j)}(u)=0$. Additionally let  $\hat{S}^{(j)}_k(t) = 1 - \hat{F}^{(j)}_k(t)$ denote the estimator for the subsurvival function $S^{(j)}_k(t)=1-F^{(j)}_k(t)$. Note that $\hat{S}^{(j)}_1$ and $\hat{K}^{(j)}$ coincide in the case $\alpha^{(j)}_2\equiv0$. But in general each CIF depends on both cause-specific hazard rates because of \eqref{distT}. 

Inference for $F_1$ is typically based on the Aalen-Johansen processes
\begin{equation*}
	W^{(j)}_{n_j}(t)= \sqrt{n_j}(\hat{F}^{(j)}_1(t)- F^{(j)}_1(t)), \quad j=1,2.
\end{equation*}
 Based on $n=n_1+n_2$ independent observations from two separate groups, we want to infer whether the CIFs for cause 1 of the two groups coincide. This results in the unpaired two-sample testing problem:
 \begin{equation} \label{Hypothesis}
 	\begin{aligned}
 	 	H_0&: F^{(1)}_1(t) = 	F^{(2)}_1(t) \text{ for all } t\in [t_1, t_2] \text{\quad versus \quad}
 	H_1: F^{(1)}_1(t) \ne 	F^{(2)}_1(t) \text{ for some }  t\in [t_1, t_2],
 	\end{aligned}
\end{equation}
for some non-empty time interval $[t_1,t_2] \subset [0, \tau_0)$ of interest. Here $\tau_0$ denotes a terminal time. If right censoring is present, it is given by $\tau_0=\min(\Tilde{\tau}^{(1)}, \Tilde{\tau}^{(2)}, \tau_C^{(1)}, \tau_C^{(2)})$ with $\Tilde{\tau}^{(j)}= \sup \{u:\int_{0}^{u}(\alpha^{(j)}_1(s) +\alpha^{(j)}_1(s)) ds)<\infty \}, j=1,2$ and $\tau_C^{(j)} \in (0,\infty], j=1,2$ is the right boundary of the (possibly group dependent) censoring time distribution. This restriction ensures that $P(X^{(j)}_i > \cdot)$ is positive on the whole interval $[0,\tau_0)$.
We also consider the usual assumption that the probability of being at risk should be bounded away from zero asymptotically on the whole interval $[t_1, t_2]$, i.e. for each group there exist deterministic functions $y^{(j)}(\cdot)$ with $\inf_{u\in[t_1, t_2]}y^{(j)}(u)>0$ such that convergence in probability holds
\begin{equation} \label{yassumption}
    \sup_{u\in[t_1, t_2]}\left|\frac{Y^{(j)}(u)}{n_j}-y^{(j)}(u)\right| \xrightarrow{p} 0 \text{ as } n_j \to \infty.
\end{equation}
In this case, the group specific Aalen-Johansen estimators are consistent and obey a functional limit theorem \citep[§IV.4.2]{andersen1993statistical}. This assumption is fulfilled in many practical applications, especially in the case of random and groupwise i.i.d. left-truncation and right censoring, if we additionally assume that $P(L_1^{(j)}\leq0)>0$, cf. \citet[Example IV.4.7]{andersen1993statistical} and \citet{wellek1990nonparametric}. Additionally, we suppose that no group size vanishes asymptotically (note that the observable sample size is random due to left-truncation), i.e.,
\begin{equation} \label{groupassumption}
    \frac{n_1}{n} \xrightarrow{p} \kappa^{(1)} \in (0,1),\quad \frac{n_2}{n} \xrightarrow{p} \kappa^{(2)} = 1 - \kappa^{(1)} \quad \text{ as } n \to \infty.
\end{equation}
As a consequence, we obtain a joint weak convergence result for group size adjusted Aalen-Johansen processes.
\begin{theorem} \label{jointconvergence}
    Suppose \eqref{yassumption} and \eqref{groupassumption} hold. Then, as $n \to \infty$, we have convergence in distribution
\begin{equation*}
    (V^{(1)}_{n_1}, V^{(2)}_{n_2}) = \left(\frac{\sqrt{n}}{\sqrt{n_1}}W^{(1)}_{n_1}, 
    \frac{\sqrt{n}}{\sqrt{n_2}}W^{(2)}_{n_2}\right) \xrightarrow{d} (\mathbb{G}^{(1)},\mathbb{G}^{(2)})
\end{equation*}
    on the Skorokhod space $D([t_1,t_2]) \times D([t_1,t_2])$ equipped with the usual Skorokhod topology where $\mathbb{G}^{(j)}, j=1,2$ are independent, mean-zero Gaussian processes with almost surely continuous sample paths and covariance functions
    \begin{equation} \label{covfunction}
    \begin{split} 
    \Gamma^{(j)}_{\mathbb{G}}(s,t) = & \frac{1}{\kappa^{(j)}}\Bigg(\int_{0}^{s \wedge t}\frac{(S_2^{(j)}(u)-F_1^{(j)}(t))(S_2^{(j)}(u)-F_1^{(j)}(s))}{y^{(j)}(u)}\alpha^{(j)}_1(u)du
    \\
        &+ \int_{0}^{s \wedge t}\frac{(F_1^{(j)}(u)-F_1^{(j)}(t))(F_1^{(j)}(u)-F_1^{(j)}(s))}{y^{(j)}(u)}\alpha^{(j)}_2(u)du\Bigg), ~ s,t \in [t_1,t_2].
    \end{split}
    \end{equation}
\end{theorem}
In the subsequent section, it is shown that a continuous functional of the joint process $(V^{(1)}_{n_1}, V^{(2)}_{n_2})$ can be used to construct an omnibus test for the two-sample testing problem \eqref{Hypothesis}. Because of the complicated covariance structure of the limiting process, in particular the lack of independent increments,  inference procedures are usually based on resampling or approximation techniques. To this end, we also present an asymptotically equivalent representation of $W^{(j)}_{n_j}$ based on the individual cause-specific martingales \eqref{martingale} by ($j=1,2$)
\begin{multline}\label{ajmartingale}
    W^{(j)}_{n_j}(t)= \sqrt{n_j} \sum^{n_j}_{i=1}\Bigg(\int_{0}^{t}\frac{S^{(j)}_2(u)-F^{(j)}_1(t)}{Y^{(j)}(u)}dM^{(j)}_{1,i}(u) \\ +
    \int_{0}^{t}\frac{F^{(j)}_1(u)-F^{(j)}_1(t)}{Y^{(j)}(u)}dM^{(j)}_{2,i}(u)\Bigg) + o_p(1),
\end{multline}
where $o_p(1)$ denotes a term converging to $0$ in probability.
A proof of this representation was first given by \citet{lin1997non}. The further simplified version presented here is due to \citet{dobler2017non}, see also \citet{bluhmki2018wild}. This representation can be used to derive bootstrap schemes for statistics based on the Aalen-Johansen process, as outlined in the next sections.

\section{Asymptotic Area based Test}
To test the Hypothesis \eqref{Hypothesis}, we propose to use the scaled area between the CIF curves (ABC) statistic defined by
\begin{equation} \label{tstat}
    T^{\text{ABC}}_n = \sqrt{n}\int_{t_1}^{t_2} |\hat{F}^{(1)}_1(t)-\hat{F}^{(2)}_1(t)|dt.
\end{equation}
Because the Aalen-Johansen estimators are step functions, with jumps only at observed event times, we can rewrite our statistics as sums. Therefore, let $k_n$ denote the number of observed events (cause 1 and 2) in the pooled sample $X^{(1)}_1, \dots X^{(1)}_{n_1},X^{(2)}_{1}, \dots, X^{(2)}_{n_2}$ after removing all observations outside the interval $[t_1, t_2]$. Because we assume existence of cause-specific hazards, ties only occur with probability $0$. We denote the ordered time points by $s_1 <\dots < s_{k_n}$ and also set $s_{0}=t_1$ and $s_{k_n+1}=t_2$. Then \eqref{tstat} becomes
\begin{equation*} \label{sumstatistic}
    T^{\text{ABC}}_n = \sqrt{n}\sum_{i=0}^{k_n}|\hat{F}^{(1)}_1(s_i)-\hat{F}^{(2)}_1(s_i)|(s_{i+1} - s_i).
\end{equation*}

\citet{lyu2020comparison} proposed the standardized test statistic $$ Z^{\text{ABC}} \coloneqq \dfrac{T^{\text{ABC}}_n - \widehat{E}(T^{\text{ABC}}_n)}{\sqrt{\widehat{Var}(T^{\text{ABC}}_n})}$$ for testing \eqref{Hypothesis} and postulated its asymptotic normality. The following result, however, shows that the actual limit distribution is more complex.
\begin{theorem} \label{h0limit}
    Suppose that \eqref{yassumption} and \eqref{groupassumption} hold. Then we have under $H_0:F^{(1)}_1(t) = F^{(2)}_1(t) = F_1(t)$ for all $t \in [t_1, t_2]$,  convergence in distribution
    \begin{equation*} \label{limitasy}
        T^{\text{ABC}}_n \xrightarrow{d} \int_{t_1}^{t_2} |\mathbb{U}(t)|dt =:U^{\text{ABC}} \text{ as } n\to \infty,
    \end{equation*}
where $\mathbb{U}$ is a zero-mean Gaussian process with almost surely continuous sample paths and covariance function
    \begin{equation*}
    \Gamma_{\mathbb{U}}(s,t) = \Gamma^{(1)}_{\mathbb{G}}(s,t) + \Gamma^{(2)}_{\mathbb{G}}(s,t)
    \end{equation*}
    with $\Gamma^{(j)}_V, j=1,2$ given in Equation \eqref{covfunction}.
\end{theorem}
Since the limit variable $U^{\text{ABC}}$ is positive, the support of its standardized version $(U^{\text{ABC}} - E(U^{\text{ABC}}))/\sqrt{Var(U^{\text{ABC}})}$ is bounded from below. This shows, that the asymptotic normality of  $Z^{\text{ABC}}$, postulated by \citep{lyu2020comparison}, does not hold. Let $q_\alpha$ denote the $(1-\alpha)$-quantile of the limit distribution in \eqref{limitasy}. The test 
$$
\varphi_{n, \alpha}= \mathbbm{1}(T^{\text{ABC}}_n > q_\alpha)
$$
then is of asymptotic level $\alpha \in (0,1)$ by Theorem \ref{h0limit}. Furthermore, the test is consistent against the general alternative $H_1$ from \eqref{Hypothesis}.
\begin{theorem} \label{consistency_test_eff}
Suppose the assumptions of Theorem \ref{h0limit} hold. Then
 we have under $H_1:F^{(1)}_1(t) \ne F^{(2)}_1(t)$ for some $t \in [t_1, t_2]$,
    \begin{equation*}
        T^{\text{ABC}}_n \xrightarrow{p} \infty \text{ as } n \to \infty,
    \end{equation*}
    and therefore $\lim_{n\to \infty} E_{H_1}(\varphi_{n, \alpha})=1$.

\end{theorem}

Consistency for the general alternative also holds for other distance-based tests mentioned previously. Unlike those procedures, the ABC statistic aims to detect the overall discrepancy of two CIFs on the time interval of interest. Such global differencesmay be overlooked by, e.g., Anderson-Darling-type statistics, which are aimed at detecting differences at boundaries of the interval \citep{dobler2017approximate}. Moreover, the general idea of the test, to reject the null hypothesis if the area between curves is large, allows for straightforward visualization and communication to practitioners.
As noted above, the covariance function of the limit process $\mathbb{U}$ has a complicated form, and  depends on usually unknown quantities. We therefore propose an asymptotically exact bootstrap method for estimating the unknown quantile of the limit distribution in the next section.

\section{Wild Bootstrap Test}
To approximate the limit distribution of the ABC statistic in Equation \eqref{limitasy}, we construct a wild bootstrap counterpart of the group size adjusted Aalen-Johansen processes $V^{(j)}_{n_j}, j=1,2$. The key idea is to use the martingale representation \eqref{ajmartingale} and replace the unknown CIFs by Aalen-Johansen estimators and the martingale increments with the randomly weighted increments of the observed transition process. This leads to the wild bootstrap processes
\begin{equation*} \label{bootprocess}
    \hat{V}^{(j)}_{n_j}(t)= \sqrt{n} \sum^{n_j}_{i=1}\Bigg(\int_{0}^{t}\frac{\hat{S}^{(j)}_2(u)-\hat{F}^{(j)}_1(t)}{Y^{(j)}(u)}G^{(j)}_{1,i}dN^{(j)}_{1,i}(u) \\ +
    \int_{0}^{t}\frac{\hat{F}^{(j)}_1(u)-\hat{F}^{(j)}_1(t)}{Y^{(j)}(u)}G^{(j)}_{2,i}dN^{(j)}_{2,i}(u)\Bigg)
\end{equation*}
for $j=1,2$. This resampling technique was first introduced by \citet{lin1997non} for right-censored observations and i.i.d. standard normal multipliers $G^{(j)}_{k,i}, k=1,2$, independent of the data. \citet{beyersmann2013weak} generalized this approach to independent left-truncated and right-censored data and also allowing the $G^{(j)}_{k,i}$ to be arbitrary i.i.d. zero-mean random variables with unit variance. \citet{dobler2017non} extended this resampling scheme even further, by allowing conditionally independent, data-dependent multipliers $G^{(j)}_{k,i}$ with finite fourth moments. The following theorem summarizes the most important cases for our setting, applied to the group size adjusted processes.
\begin{theorem} \label{jointcondconvergence}
    Let for each group $j=1,2$, $G^{(j)}_{k;i}$, $k=1,2,i=1,\dots,n_j$ be rowwise-i.i.d. random variables with $E(G^{(1)}_{1;1})=0, Var(G^{(1)}_{1;1})=1, E(G^{4~(1)}_{1;1}) < \infty $, independent of the data $(L^{(j)}_i,X^{(j)}_i,\delta^{(j)}_i), i = 1,...,n_j^\prime$. Suppose  \eqref{yassumption} and \eqref{groupassumption} holds. Then, as $n \to \infty$, we have convergence in distribution
    \begin{equation*}
        (\hat{V}^{(1)}_{n_1}, \hat{V}^{(2)}_{n_2})  \xrightarrow{d} (\mathbb{G}^{(1)},\mathbb{G}^{(2)})
    \end{equation*}
    on $D([t_1,t_2]) \times D([t_1,t_2])$ conditional on  the data, in probability. Here $\mathbb{G}^{(j)}, j=1,2$ are the same processes as in Theorem \ref{jointconvergence}.
\end{theorem}

To apply this result to our test statistic, note that under the null hypothesis, $0=F^{(1)}_1(t)-F^{(2)}_1(t)$, and hence Equation \eqref{tstat} can be rewritten as
\begin{equation*} \label{h0representation}
    T^{\text{ABC}}_n = \sqrt{n}\int_{t_1}^{t_2} |(\hat{F}^{(1)}_1(t) -F^{(1)}_1(t)) -(\hat{F}^{(2)}_1(t)-F^{(2)}_1(t))|dt =\int_{t_1}^{t_2} |V^{(1)}_{n_1}(t) - V^{(2)}_{n_2}(t)|dt.
\end{equation*}
Substituting the processes $V^{(1)}_{n_1}, V^{(2)}_{n_2}$ with the wild bootstrap counterparts $\hat{V}^{(1)}_{n_1}, \hat{V}^{(2)}_{n_2}$, we obtain the resampling statistic
\begin{equation*}\label{resampstat}
\begin{aligned}
    \Tilde{T}^{ABC}_n &= \int_{t_1}^{t_2} |\hat{V}^{(1)}_{n_1}(t) - \hat{V}^{(2)}_{n_2}(t)|dt
    \\
    &=\sum_{i=0}^{k_n}|\hat{V}^{(1)}_{n_1}(s_i)-\hat{V}^{(2)}_{n_2}(s_i)|(s_{i+1} - s_i).
\end{aligned}
\end{equation*}
Based on Theorem \ref{jointcondconvergence}, we immediately obtain a conditional weak convergence result.
\begin{theorem} \label{weaklimitabcboot}
    Let $\hat{V}^{(1)}_{n_1}, \hat{V}^{(2)}_{n_2}$ be wild bootstrap processes with multipliers that fulfill the conditions of Theorem \ref{jointcondconvergence}. Suppose \eqref{yassumption} and \eqref{groupassumption} holds. Then as $n \to \infty$, we have under $H_0$ as well as $H_1$, 
    \begin{equation*}
        \Tilde{T}^{\text{ABC}}_n \xrightarrow{d} U^{\text{ABC}}
    \end{equation*}
    conditionally on the data, in probability.
\end{theorem}
Thus, the wild bootstrap statistic asymptotically mimics the null distribution of $T^{\text{ABC}}_n$. Let $q^{\ast}_{n, \alpha}$ be the conditional (data dependent) $(1-\alpha)$-quantile of $\Tilde{T}^{\text{ABC}}_n$, and denote by $$\varphi^{\ast}_{n, \alpha}= \mathbbm{1}(T^{\text{ABC}}_n > q^{\ast}_{n, \alpha})$$ the wild bootstrap counterpart of the test $\varphi_{n, \alpha}$.
\begin{theorem} \label{resamplingtestproperties}
    The wild bootstrap test $\varphi^{\ast}_{n, \alpha}$ is asymptotically equivalent to the unconditional test $\varphi_{n, \alpha}$ under the null hypothesis, i.e., $\lim_{n\to \infty} E_{H_0}(|\varphi_{n, \alpha} - \varphi^{\ast}_{n, \alpha}|)=0$, and therefore asymptotically exact. Furthermore it is consistent under the general alternative $H_1$, i.e., $\lim_{n\to \infty} E_{H_1}(\varphi^{\ast}_{n, \alpha})=1$.
\end{theorem}
For implementation, let $n_j^{\ast} \in \mathbb{N}$ denote the number of all observed events in group $j$. Denote by $w^{(j)}_i, j=1,2, i=1,\dots, n_j^{\ast}$ the realizations of the bootstrap multipliers and by $u^{(j)}_i$ the observed event times. Generating only one multiplier per observation is sufficient, because for each observation $\Delta N^{(1)}_{k,i}(\infty) +\Delta N^{(2)}_{k,i}(\infty) \in \{0,1\}$ holds. Furthermore a censored observation does not contribute to the wild bootstrap process, and no multiplier has to be generated in this case. One realization of the wild bootstrap process $\hat{V}^{(j)}_{n_j}$ is then given by
\begin{equation*}
    \hat{V}^{\ast, (j)}_{n_j}(t)= \sqrt{n}\sum_{i=1}^{n_j^{\ast}}w^{(j)}_i\frac{\hat{S}^{(j)}_2(u^{(j)}_i)\Delta N^{(j)}_{1,i}(u^{(j)}_i)+ \hat{F}^{(j)}_1(u^{(j)}_i)\Delta N^{(j)}_{2,i}(u^{(j)}_i) - \hat{F}^{(1)}_1(t)}{Y^{(j)}(u^{(j)}_i)},
\end{equation*}
where $\Delta N^{(j)}_{k,i}(u^{(j)}_i)=1$ if a cause $k$ event was observed for $i$ at $u_i$ and $\Delta N^{(j)}_{k,i}(u^{(j)}_i)=0$ otherwise. $\hat{V}^{\ast, (j)}_{n_j}$ is a step function, with jumps only at event times, see Section 5 of \citet{beyersmann2013weak}. Inserting this representation into \eqref{resampstat}, we obtain a wild bootstrap replicate of the ABC statistic in the following algorithm, which can be used to implement the wild bootstrap test.
\begin{algorithm} Let $\alpha \in (0,1)$ be the significance level, and $B \in \mathbb{N}$ a sufficiently large (e.g. $B=1000$) number of wild bootstrap replications.
    \begin{enumerate}[label=(\arabic*)]
        \item Compute the test statistic $T^{\text{ABC}}_n$ in Equation \eqref{tstat}.
        \item Generate a random variable that fulfills the conditions of Theorem \ref{jointcondconvergence} for every observation with an observed event (cause 1 or 2).
        \item Compute the wild bootstrap statistic $\Tilde{T}^{\text{ABC}}_n$ in Equation \eqref{resampstat}.
        \item Repeat steps (2) and (3) $B$ times.
        \item Compute the empirical $(1-\alpha)$ quantile $q^{\ast}_{n, \alpha, B}$ of the collection of wild bootstrap statistics $\Tilde{T}^{\text{ABC}}_{n,1}, \dots \Tilde{T}^{\text{ABC}}_{n,B}$.
        \item Reject the null Hypothesis $H_0$ \eqref{Hypothesis} if and only if $T^{\text{ABC}}_n > q^{\ast}_{n, \alpha, B}$.
    \end{enumerate}
\end{algorithm}
This algorithm also highlights an additional advantage of the wild bootstrap approach: Since only the test statistic itself is reweighted, the method is computationally  efficient. This contrasts other resampling approaches such as the Efron bootstrap \citep{efronBoot1979} which requires resampling the original data and a re-computation of all necessary quantities in the test statistic for each resampling step, which may be prohibitive for large datasets.

\begin{remark}\label{rem:cifjumps}
    All results of the previous sections can be extended to include situations where the cause-specific hazard functions \eqref{hazards} do not exist, i.e., the CIFs are not continuous, if we exclude the right endpoint $t_2$ in our hypothesis \eqref{Hypothesis} and assume that the CIFs have only finitely many jumps. However, the covariance function of the Gaussian process $\mathbb{U}$ changes and the wild bootstrap processes have to be adjusted accordingly \citep{dobler2017discontinuity, dobler2024erratum}. More details are provided in Appendix B.
\end{remark}

\section{Simulations}
This section reports a simulation study assessing the finite-sample performance of the proposed ABC test in comparison with existing procedures. Section~\ref{subsec:simsetup} describes the data-generating mechanisms and implementation details. Section~5.2 summarizes results on type I error control and power.

\subsection{Simulation Setup}\label{subsec:simsetup}
To compare our area based wild bootstrap test with the standardized area test proposed by \citet{lyu2020comparison} as well as other tests for equality of CIFs, we conducted a simulation study. The following tests were included:
\begin{enumerate}[label=(\arabic*)]
    \item Our area based wild bootstrap test with independent standard normal, centered unit Poisson, and Rademacher multipliers, subsequently denoted by $\varphi_{\text{N}}^{\text{ABC}}$, $\varphi_{\text{Po}}^{\text{ABC}}$ and $\varphi_{\text{R}}^{\text{ABC}}$.
    \item The asymptotic test $\varphi_{\text{Z}}^{\text{ABC}}$ of \citet{lyu2020comparison} based on the standardized test statistic $Z^{\text{ABC}}$ and the incorrect standard normal approximation.
    \item A variant of the asymptotic test of \citet{pepe1991inference}, which is based on the integrated CIF difference $\int_{t_1}^{t_2}(\hat{F}^{(1)}_1(t)-\hat{F}^{(2)}_1(t))dt$ denoted by $\varphi^{\text{Pepe}}$. This test is a special case of the tests of \citet{munko2025rmtl} and we use the variance estimator proposed therein for studentization.
    \item Wild bootstrap versions of the (unweighted) Kolmogorov-Smirnov and Cramér-von-Mises-type tests, using  the same multipliers as in (1) , denoted by $\varphi_{\text{N}}^{\text{KS}}$, $\varphi_{\text{Po}}^{\text{KS}}$, $\varphi_{\text{R}}^{\text{KS}}$, $\varphi_{\text{N}}^{\text{CvM}}$, $\varphi_{\text{Po}}^{\text{CvM}}$ and $\varphi_{\text{R}}^{\text{CvM}}$.
    \item We additionaly adopted Remark 2 of \citet{dobler2017non} and included variants of all bootstrap tests with slightly modified multipliers $\Tilde{G}^{(j)}_{k;i}= G^{(j)}_{k;i}(1+r_n)$ with $$r_n=\frac{n_1+n_2}{n_1n_2}.$$ This correction does not change the asymptotic behavior of the tests, and is aimed at improving type I error control in smaller samples, which can be challenging for wild bootstrap tests in some scenarios \citep{dobler2014bootstrapping}.
    We denote the correction versions by adding a prefix "c" to the test index. 
    Accordingly, we denote the corrected ABC tests by $\varphi_{\text{cN}}^{\text{ABC}}$, $\varphi_{\text{cPo}}^{\text{ABC}}$, and $\varphi_{\text{cR}}^{\text{ABC}}$, respectively. The corrected versions for the Kolmogorov-Smirnov and Cramér-von Mises type wild bootstrap tests are denoted similarly.
\end{enumerate}
The Kolmogorov-Smirov-type test with normal multipliers was introduced by \citet{lin1997non}. \citet{dobler2017non} demonstrated the validity of wild bootstrap Kolmogorov-Smirnov and Cramér-von-Mises-type tests for multipliers that fulfill the conditions of Theorem \ref{jointcondconvergence}. It is also worth noting that $\varphi^{\text{Pepe}}$ is not an omnibus test for our general hypothesis \eqref{Hypothesis}. We nevertheless include it because it is among the most well-known tests in competing risks analyses and was the first hypothesis test for competing risks based directly on the CIFs.

All simulations were carried out utilizing the R-computing environment, version 4.5.0 \citep{Rsoftware}, with $N_{\text{sim}}=5000$ simulation runs. Moreover for the resampling tests, $B=1000$ bootstrap runs in each simulation step were used. For simulating the competing risk event times, the method of \citet{beyersmann2011competing} was used. The R Code implementing this method, as well as the wild bootstrap resampling scheme, was provided by Dennis Dobler. 
The following data-generating mechanisms were used:
\begin{enumerate}[label=(\arabic*)]
    \item The setup of \citet{dobler2014bootstrapping}. This model is specified by the cause-specific hazard functions of the first group as 
    \[\alpha_1^{(1)}(t)= \exp(-t) \text{ and } \alpha_2^{(1)}(t)= (1-\exp(-t))\]
    and for the second group as
    \[\alpha_1^{(2)}(t) \equiv c \equiv 2 - \alpha_2^{(2)}(t)\]
    with $c \in [0,1]$. Here $c=1$ corresponds to the null hypothesis of equal CIFs of the first cause. It is also worth noting, that even under null hypothesis, the cause-specific hazard rates for the two groups are different. For $c<1$ the alternative is true. We have chosen $c=0.65$ for our simulations. The time interval of interest was set to $[t_1,t_2]=[0,1.5].$ This setup leads to an ordered alternative, $F_1^{(1)}(t) > F_1^{(2)}(t)$ for $t \in (0,1.5]$.
    \item Model 3 of \citet{bajorunaite2008comparison}, given by the CIFs for both groups as
    \[F_1^{(j)}(t) = p(1-\exp(-\beta^{(j)}t^{a^{(j)}}))\]
    and
    \[F_2^{(j)}(t) = (1-p)(1-\exp(-\beta^{(j)}t^{a^{(j)}}))\]
    for $p\in(0,1)$, $\beta^{(j)}, a^{(j)} >0$ and $j=1,2$. The parameter $p$ specifies the probability of a cause 1 event. Notice that the CIFs in this case are scaled Weibull distribution functions. For $\beta^{(1)}=\beta^{(2)}$ and $a^{(1)}=a^{(2)}$ the null hypothesis is true; otherwise 
    the alternative is implied. The time interval of interest was again set to $[t_1,t_2]=[0,1.5]$. For simulations under the null hypothesis the parameters were chosen as $\beta=\frac{1}{3^{1.2}}$ and $a=3$. For the alternative the parameters were chosen such that the CIFs in the two groups for cause 1 cross, and the integral $\int_{0}^{1.5} F_1^{(1)}(t)- F_1^{(2)}(t) dt$ is close to zero. This was achieved with the parameters $\beta^{(1)}=\frac{1}{3.8^{1.2}},\beta^{(2)}=\frac{1}{1.5^{1.5}}, a^{(1)}=3.8, a^{(2)}=1.5$. Furthermore, the probability of a cause 1 event was set to $p=0.6$ for both the null hypothesis and the alternative.
\end{enumerate}
As sample sizes $(n_1,n_2)=(50, 50), (50, 100), (100, 400), (100,100), (200,200), (400,400)$ were considered.
These sample sizes were chosen to reflect small, moderate, and large sample scenarios, including both balanced and unbalanced group sizes. Each setting was simulated with and without right-censoring, where censoring times were simulated as independent exponential random variables for the first group with rate parameters $\lambda = 1/3, 2/3, 1$ leading to light, medium and heavy censoring, and as independent uniform random variables on $[0,b]$ with upper bounds $b = 1.6, 2.5$ leading to medium and heavy censoring. Here, we considered all possible parameter combinations for the censoring distributions, resulting in six different censoring setups additionally to the uncensored case. For our first simulation setup, equal censoring distributions do not lead to equal censoring rates, because even though the CIFs corresponding to the first risk are equal under the null hypothesis, the survival functions in the two groups are different. The resulting censoring rates range from $20\%$ to $68\%$, the detailed rates are presented in Table~\ref{tab:censrates} in the Appendix. The nominal significance level was set to $\alpha=0.05$ for all settings. Left-truncation was not included in the simulations, as not all considered tests are designed to handle left-truncated data.

\subsection{Results}
For clarity, we only present and analyze the results of Model~2 in detail in this section. The results of Model~1 can be found in the Appendix~\ref{ssec:model1}.

Figure \ref{fig:bysample_mod2_h0} displays the results for Model 2 under the null hypothesis stratified by sample sizes (additional stratified results are available in Appendix \ref{ssec: add plots}). For the wild bootstrap tests, we only present the results based on Poisson multipliers as the behavior under normal and Rademacher multipliers is very similar.
Overall, the asymptotic, wild bootstrap, and Pepe tests are in general too liberal for smaller sample sizes.
Moreover, Figure~\ref{fig:bysample_mod2_h0} clearly shows that $\varphi_{\text{Z}}^{\text{ABC}}$ is not mathematically valid. While the Type I error of all wild bootstrap tests approaches the correct significance level as the sample size increases, $\varphi_{\text{Z}}^{\text{ABC}}$ becomes increasingly conservative for large samples, even though the Type I error for sample sizes of $n_1 = n_2 = 50$ corresponds approximately to the specified significance level.
\begin{figure}[h!]
    \centering
    \includegraphics[width=\linewidth]{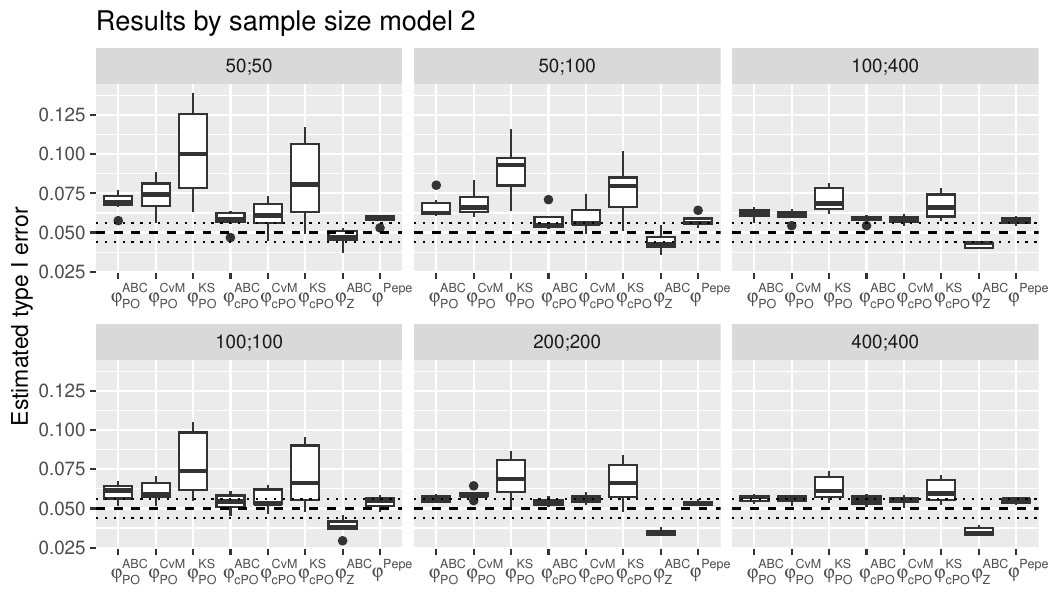}
    \caption{Estimated type I error under Model~2 stratified by sample sizes}
    \label{fig:bysample_mod2_h0}
\end{figure}

\begin{figure}[h!]
    \centering
    \includegraphics[width=\linewidth]{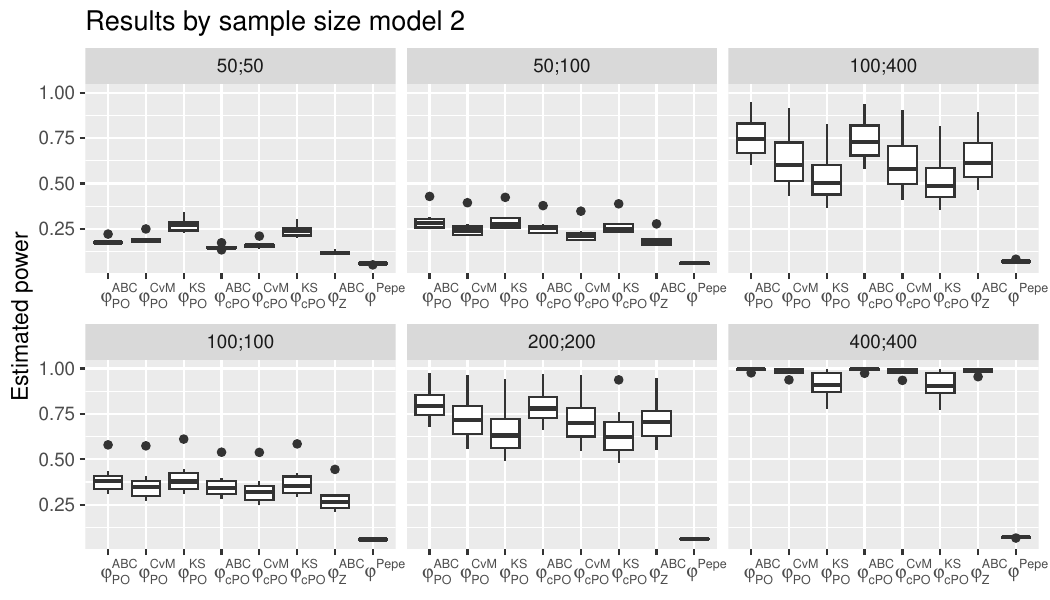}
    \caption{Estimated power under Model 2 stratified by sample sizes}
    \label{fig:bysample_mod2_h1}
\end{figure}

This behavior results in a loss of power under the alternative as illustrated in Figure~\ref{fig:bysample_mod2_h1}, especially for small to medium sample sizes (up to 100 individuals per group). The correction factor $r_n$ substantially improves type-I error control of the wild bootstrap approach in smaller samples, without large impacts on power. By construction, $\varphi^{\text{Pepe}}$ can not distinguish $H_0$ from $H_1$ which is clearly visible in the simulation results and demonstrates that this test is not suitable for crossing CIFs.

For model 1, the observations are similar than for model 2, with the exception that $\varphi^{\text{Pepe}}$ has comparable power to the other tests, which was expected as the test is consistent for ordered alternatives \citep{pepe1991inference}.

\section{Real-Data Application}
To illustrate the proposed area based test in practice and to compare it with the standardized statistic and the other procedures considered in the simulations, we analyze the data of 2279 individuals that received a bone marrow transplant at the European Society for Blood and Marrow Transplantation (EBMT) between 1985 and 1998, see \citet{fiocco2008reduced} for a more detailed description of the dataset. The data is openly available trough the R package \texttt{mstate} \citep{mstate2011}. The event of interest is death after transplantation, whereas remission is the competing event. Similar to the analysis of \citet{lyu2020comparison}, we focus on the comparison of two groups: 545 patients without a gender matching donor (gender mismatch group), and 1734 with a gender matching donor (no gender mismatch group). Both groups are subject to right-censoring, with censoring rates of $61\%$ and $57\%$. The event of interest was observed for $388~(22\%)$ and $145~(27\%)$ patients, respectively. To examine the effects of a donor mismatch for different age groups (under 20 years, 20-40 years and over 40 years), an additional subgroup analysis was performed.
\begin{figure}[h]
\caption{Estimated cumulative incidence for death after bone marrow transplantation for the different age and gender matching groups.}
\includegraphics[width=\textwidth]{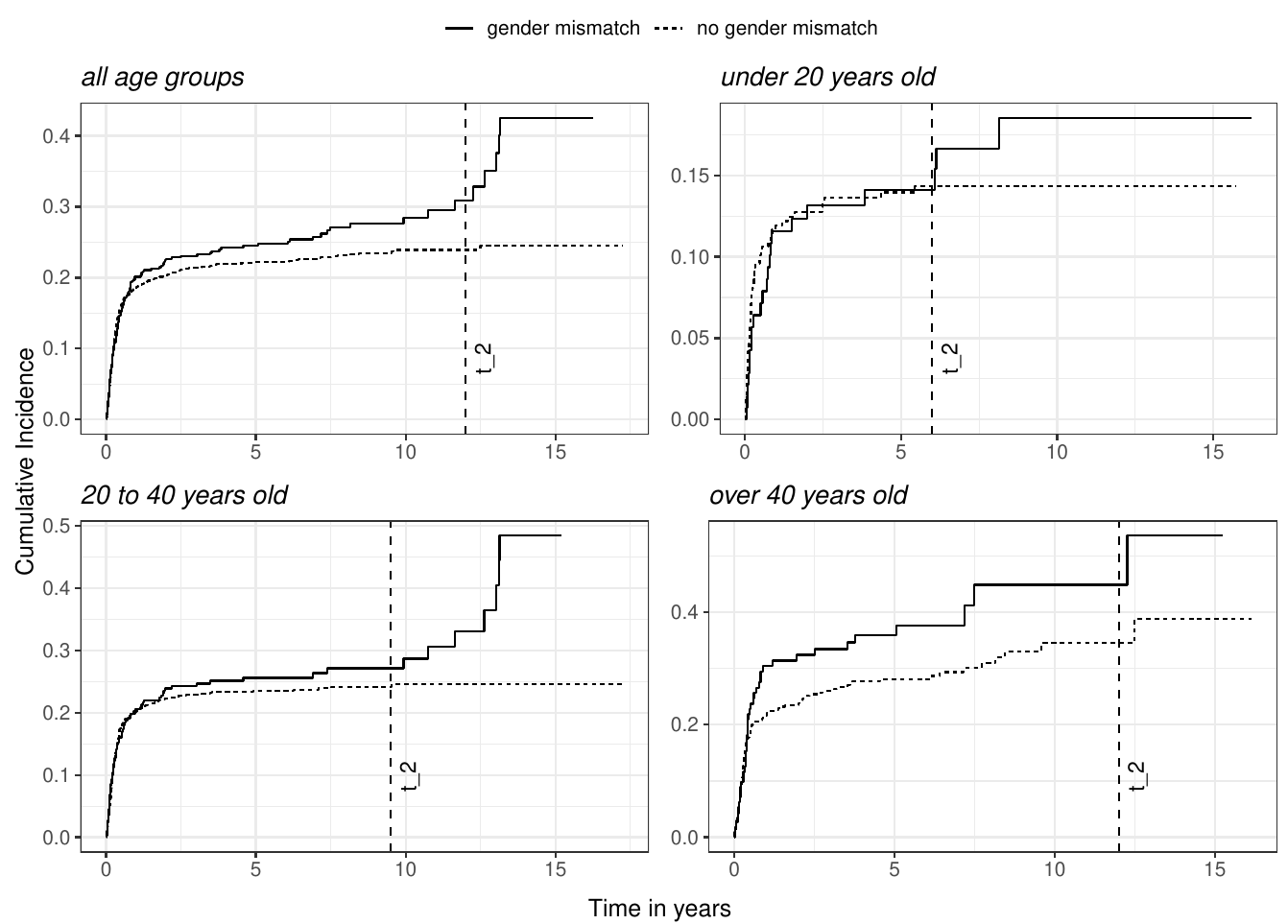}
\label{fig:example1}
\end{figure}
Figure~\ref{fig:example1} reveals a clear difference between the Aalen-Johansen estimators for the subset of over 40 year old patients throughout the whole time period, whereas a noticeable difference only emerges late in the study for the other subgroups as well as the whole sample.

The time intervals for the analysis were chosen such that at least one event was observed for the match and mismatch group after the endpoint. This led to the time intervals $[0, 12]$ (all age groups and over 40 years), $[0,9.5]$ (between 20 and 40 years) and $[0,6]$ (under 20 years). For the resampling based tests, $B=1000$ bootstrap resp. permutation steps were used. For the wild bootstrap tests, we chose modified centered Poisson weights $G^{(j)}_{k;i}(1+r_n), G^{(j)}_{k;i}\sim \text{Po}(1)-1$ and $r_n$ as defined in Section~\ref{subsec:simsetup}. Because most of the considered tests assume existing cause-specific hazard rates, we added a mean zero uniform random variable with very small support to each observed time in order to break any ties present in the original dataset. The $P$-values of all tests considered in the simulations are presented in Table \ref{data_ex table}.

\begingroup

\renewcommand{\arraystretch}{1.5} 
\begin{table}[tbh]
\centering
\caption{$P$-values of different tests for the null hypothesis of equal CIFs for gender match and mismatch individuals for the EBMT data example. Values below the level of 5\% are printed in bold.}
\begin{tabular}{|r|llll|}
  \hline
Group & all age groups & under 20 years old & 20 to 40 years old & over 40 years old \\
  \hline
  $n_1, n_2$ & $545, 1734$ & $143, 408$ & $290, 923$ & $112, 403$ \\
  $[t_1,t_2]$ & [0,12] & [0,6] & [0,9.5] & [0,12] \\\hline
  $\varphi_{\text{Pois}}^{\text{ABC}}$ & 0.142 & 0.926 & 0.455 & \textbf{0.037} \\ 
  $\varphi_{\text{Pois}}^{\text{CvM}}$ & 0.120 & 0.867 & 0.435 & 0.055 \\ 
  $\varphi_{\text{Pois}}^{\text{KS}}$ & \textbf{0.021} & 0.544 & 0.470 & \textbf{0.040} \\ 
  $\varphi_{\text{Z}}^{\text{ABC}}$ & 0.178 & 0.161 & 0.705 & \textbf{0.048} \\ 
  $\varphi^{\text{Pepe}}$ & 0.146 & 0.888 & 0.519 & 0.076 \\ 
   \hline
\end{tabular}
\label{data_ex table}
\end{table}

\endgroup

For the full dataset, only the Kolmogorov-Smirnov type test showed significance at the 5\% level. Since this test showed a very liberal behaviour in our simulations, this finding should be interpreted with caution. For the subgroup analyses, significant results were found only for the group of over 40-year-old patients. Here, our area based test, the Kolmogorov-Smirnov type test and the asymptotic area based test of \citet{lyu2020comparison} indicate significant differences between groups. As the example is merely illustrative, no multiplicity adjustment was performed.

\FloatBarrier
\section{Conclusion and Discussion}
We studied the two-sample testing problem of equality of cumulative incidence functions (CIFs) from two independent groups, with a specific focus on settings with crossing CIFs. Under minimal assumptions, allowing for left-truncated and right-censored observations, we 
revisited the area-based ABC test proposed by 
\citet{lyu2020comparison} and proved that its claimed asymptotic normality is not true. We derived the correct limit distribution of this area between the CIF curves statistic. As it depends on unknown quantities,  we proposed a flexible wild bootstrap approach that yields an asymptotically exact and consistent test for general alternatives.

Our simulation study demonstrated that
the test proposed by \citet{lyu2020comparison} exhibits substantial deviations from the nominal level, in line with the theoretical findings. In comparison, the proposed wild bootstrap ABC test performs comparably to established tests in case of ordered alternatives, and superior for crossing CIFs in many scenarios.

Beyond its statistical properties, the proposed ABC test also has practical advantages: The statistic admits an intuitive geometric interpretation as the area between CIF curves. This facilitates visualization and communication of results. Moreover, our wild bootstrap implementation is computationally efficient, as it relies on reweighting observed events rather than repeated bootstrapping from the sample. 

Several directions for future research are of interest. First, further numerical investigation of the performance of the proposed methods in settings with discontinuous CIFs or with left-truncated data would be interesting. Second, the geometric intuition of the ABC statistic suggests the possibility of constructing a similarity or distance measure based on the area between curves. Such measures could be useful beyond hypothesis testing and the mathematical details will be explored in a forthcoming paper.

\section*{Acknowledgments}
Marc Ditzhaus, Merle Munko, and Markus Pauly gratefully acknowledge support from the Deutsche Forschungsgemeinschaft (Grant No. DI 2906/1-2 (Ditzhaus, Munko, Pauly) and GRK 2297 MathCoRe (Ditzhaus, Munko)).
Simon Mack would like to thank his previous affiliation Otto von Guericke University Magdeburg where part of the work was done.
We would also like to thank Dennis Dobler for helpful discussions and for providing an implementation of the wild bootstrap resampling scheme.
\newpage
\appendix
\section*{Appendix}

The appendices are organized as follows: Appendix A contains the proofs of the results from the main part, as well as additional mathematical results that are necessary for these proofs. Appendix B contains some details regarding Remark \ref{rem:cifjumps}. In Appendix C we present the remaining simulation results from the simulation study of the main paper.

\section{Proofs}\label{appendix:proofs}
In this section, the proofs of all stated results are presented. Here and throughout let $ \mathcal{T} = [0, \tau] \subset \mathbb{R}$ be a closed, finite and non-empty interval. Denote by $D(\mathcal{T})$ the Skorokhod space of real valued càdlàg (right-continuous with existing left-hand limits) functions on $\mathcal{T}$ which we endow with the topology induced by the Skorokhod metric $d_S$ as defined, e.g. in \citet{billingsley1999convergence}. Furthermore, denote by $d_{\infty}$ the metric induced by the uniform norm. For the sake of completeness, we summarize some key properties of càdlàg functions and their convergence.

\begin{lemma} \label{lem3}
	Let $f, f_n \in D(\mathcal{T})$ for all $n\in \mathbb{N}$.
	 Denote the set of all continuity points of $f$ by $C(f) \coloneqq \{t \in \mathcal{T} \:: f \: \text{is continuous in} \: t\}.$
	\begin{enumerate}[label = (\alph*)]
		\item $f$ has at most countably many discontinuities,
		\item $f$ is bounded: $\sup_{t \in \mathcal{T}} |f(t)| < \infty$,
		\item $f$ is Borel measurable.
        \item $f_n \xrightarrow{d_{\infty}} f \implies f_n \xrightarrow{d_S} f$ as $n \to \infty$.
		\item $f_n \xrightarrow{d_S} f \implies f_n(t) \to f(t)$ for all $ t \in C(f)$ as $n \to \infty$.
		\item If $f_n \xrightarrow{d_S} f$ as $n \to \infty$, then $f_n$ is uniformly bounded.
	\end{enumerate}
\end{lemma}
\begin{proof}
	For a proof of (a)-(e) see Section 12 in \citet{billingsley1999convergence},
	 (f) follows immediately from the definition of the Skorokhod metric.
\end{proof}

As an application of this lemma, we establish an important connection between $D(\mathcal{T})$ and the $L^p$ spaces.
\begin{lemma} \label{lem4}
	Let $f, f_n \in D(\mathcal{T})$ for all $n \in \mathbb{N}$ and denote by $L^p(\mathcal{T}) \coloneqq L^p(\mathcal{T}, \mathcal{B}(\mathcal{T}), \lebesgue), 1\le p < \infty$, the space of $p$-times integrable functions with respect to the Lebesgue measure $\lebesgue$ on $\mathcal{T}$. Furthermore, let $d_{L^p}$ be the metric induced by the $L^p$-norm. Then  $f_n, f \in L^p(\mathcal{T})$ for all $n \in \mathbb{N}$ and
	\[ f_n \to f \: \text{in} \: (D(\mathcal{T}), d_S) \text{ as } n \to \infty \text{ implies } f_n \to f \: \text{in} \: (L^p(\mathcal{T}), d_{L^p}) \text{ as } n \to \infty. \]
\end{lemma}
\begin{proof}
	First, Lemma \ref{lem3} (b)+(c) implies $D(\mathcal{T}) \subset L^p(\mathcal{T})$, and therefore $f_n, f \in L^p(\mathcal{T})$ for all $n \in \mathbb{N}$. Now assume $f_n \xrightarrow{d_S} f$ as $n \to \infty$. Because countable sets have Lebesgue measure $0$, Lemma \ref{lem3} (a) + (e) yields $f_n \to f \: \lebesgue$ - almost everywhere. Furthermore, the sequence $(f_n)$ can be uniformly bounded by a constant $M \in \mathbb{R}$ by Lemma \ref{lem3} (f). Therefore, the dominated convergence theorem implies $f_n \xrightarrow{d_{L^p}} f$ as $n \to \infty$.
\end{proof}
As a corollary, we obtain the continuity of an important functional.
\begin{corollary} \label{contfunctional}
	The mapping $\Psi:D(\mathcal{T}) \times D(\mathcal{T}) \to \mathbb{R}$ defined by $\Psi(f,g) \coloneqq \int_{0}^{\tau} |f(t)-g(t)| dt$, $\tau \in \mathcal{T}$ is continuous.
\end{corollary}
With these preliminary results, we can prove the results of the main paper.

\begin{proof}[Proof of Theorem \ref{jointconvergence}]
    The weak convergence of the individual processes $W^{(j)}_{n_j}, j=1,2$ is a special case of Theorem IV.4.2. in \citet{andersen1993statistical}. Therefore the joint convergence follows from the assumed independence and the non-vanishing groups assumption \eqref{groupassumption} in combination with Slutsky's lemma.
\end{proof}

\begin{proof}[Proof of Theorem \ref{h0limit}]
    By Corollary \ref{contfunctional} the functional $\Psi$ is continuous. Therefore, the map $\Psi_{|[t_1,t_2]^2}: D([t_1,t_2]) \times D([t_1,t_2]) \to \mathbb R, \Psi_{|[t_1,t_2]^2}(f,g) := \int_{t_1}^{t_2} |f(t)-g(t)| dt$ is continuous as well. Now we can rewrite the ABC statistic under $H_0$ in terms of $\Psi_{|[t_1,t_2]^2}$ and the processes $V^{(j)}_{n_j}(t), j=1,2$ as follows, see also Equation \eqref{h0representation},
    \begin{equation*}
        T^{\text{ABC}}_n = \sqrt{n}\int_{t_1}^{t_2} |\hat{F}^{(1)}_1(t)-\hat{F}^{(2)}_1(t)|dt = \int_{t_1}^{t_2} |V^{(1)}_{n_1}(t)-V^{(2)}_{n_2}(t)|dt = \Psi_{|[t_1,t_2]^2}(V^{(1)}_{n_1},V^{(2)}_{n_2}).
    \end{equation*}
    The result now follows from Theorem \ref{jointconvergence}, the continuous mapping theorem, and the distributional equality $\mathbb{U}\stackrel{d}{=}\mathbb{G}^{(1)}-\mathbb{G}^{(2)}$ since $\mathbb{G}^{(1)}$ and $\mathbb{G}^{(2)}$ are independent Gaussian processes.
\end{proof}

\begin{proof}[Proof of Theorem \ref{consistency_test_eff}]
    Due to Assumption \eqref{yassumption}, the Aalen-Johansen estimators are uniformly consistent on the whole interval $[t_1,t_2]$, i.e., 
    \begin{equation*}
      \sup_{t\in[t_1, t_2]}  |\hat{F}^{(j)}_1(t) - F^{(j)}_1(t)| \xrightarrow{p} 0,\quad j=1,2
    \end{equation*}
    as $n \to \infty$, see \citet{andersen1993statistical} Theorem IV.4.1. By Lemma \ref{lem3} (d), uniform convergence implies convergence in the Skorokhod topology. Therefore we have as $n \to \infty$,
    \begin{equation*}
        (\hat{F}^{(1)}_1, \hat{F}^{(2)}_1) \xrightarrow{p} (F^{(1)}_1, F^{(2)}_1)
    \end{equation*}
    on $D([t_1,t_2]) \times D([t_1,t_2])$ as well. By applying the functional  $\Psi_{|[t_1,t_2]^2}$, we get
    \begin{equation} \label{functionalconsistency}
        \Psi_{|[t_1,t_2]^2}(\hat{F}^{(1)}_1, \hat{F}^{(2)}_1) \xrightarrow{p} \Psi_{|[t_1,t_2]^2}(F^{(1)}_1, F^{(2)}_1) = \int_{t_1}^{t_2} |F^{(1)}_1(t)-F^{(2)}_1(t)|dt
    \end{equation}
    as $n \to \infty$ by the continuous mapping theorem. Now suppose $F^{(1)}_1(\Tilde{t})\ne F^{(2)}_1(\Tilde{t})$ for some $\Tilde{t} \in [t_1,t_2)$. Without loss of generality suppose $F^{(1)}_1(\Tilde{t}) > F^{(2)}_1(\Tilde{t})$. Because CIFs are right-continuous there exists some small $\varepsilon\in (0, t_2 - \Tilde t)$ such that $F^{(1)}_1(t) > F^{(2)}_1(t)$ 
    for all $t \in [\Tilde{t}, \Tilde{t} +\varepsilon]\subset [t_1, t_2]$, 
    which implies
    \begin{equation*}
          \int_{t_1}^{t_2} |F^{(1)}_1(t)-F^{(2)}_1(t)|dt \ge \int_{\Tilde t}^{\Tilde t + \varepsilon} |F^{(1)}_1(t)-F^{(2)}_1(t)|dt >0.
    \end{equation*}
    By combining this result with \eqref{functionalconsistency}, we obtain
    \begin{equation*}
        T_n^{\text{ABC}}= \sqrt{n}\Psi_{|[t_1,t_2]^2}(\hat{F}^{(1)}_1, \hat{F}^{(2)}_1) \xrightarrow{p} \infty \text{ as } n \to \infty
    \end{equation*}
    by Slutsky´s lemma. Note that $\Tilde{t} \in [t_1,t_2)$ is no restriction, because our assumption of existing cause-specific hazard rates implies that the CIFs are continuous. Therefore $F^{(1)}_1(t_2)\ne F^{(2)}_1(t_2)$ and $F^{(1)}_1(t) = F^{(2)}_1(t)$ for all $t \in [t_1, t_2)$ is not possible.
\end{proof}

\begin{proof}[Proof of Theorem \ref{jointcondconvergence}]
    The conditional weak convergence of one-sample wild bootstrap processes was proven by \citet{dobler2017non} in their Theorem 1. The result now follows by the same argument as in the Proof of Theorem \ref{jointconvergence}.
\end{proof}

\begin{proof}[Proof of Theorem \ref{weaklimitabcboot}]
    By Theorem \ref{jointcondconvergence} the wild bootstrap processes, given the data, mimic asymptotically the distribution of the scaled Aalen-Johansen processes $(V^{(1)}_{n_1},V^{(2)}_{n_2})$. The resampling statistic can be expressed trough the functional $\Psi_{|[t_1,t_2]^2}$ and the wild bootstrap processes as
    \begin{equation*}
        \Tilde{T}^{\text{ABC}}_n = \int_{t_1}^{t_2} |\hat{V}^{(1)}_{n_1}(t) - \hat{V}^{(2)}_{n_2}(t)|dt = \Psi_{|[t_1,t_2]^2}(\hat{V}^{(1)}_{n_1}, \hat{V}^{(2)}_{n_2}).
    \end{equation*}
    Therefore the result follows from the continuous mapping theorem.
\end{proof}

\begin{proof}[Proof of Theorem  \ref{resamplingtestproperties}]
    Under $H_0$, the limit distributions of the original statistic $T^{\text{ABC}}_n$ and its resampling counterpart $\Tilde{T}^{ABC}_n$ coincide by Theorem \ref{weaklimitabcboot}. As this limit distribution (i.e. the distribution of $U^{ABC}$) is continuous, the asymptotic equivalence of unconditional and wild bootstrap test under $H_0$ follows from Lemma 1 of \citet{janssen2003bootstrap}. Because the resampling statistic, given the data, converges to the same, real valued limit under $H_1$, it is tight especially. The consistency of the wild bootstrap now follows from Theorem 7 of \citet{janssen2003bootstrap}, because $T^{\text{ABC}}_n$ diverges under $H_1$ by Theorem \ref{consistency_test_eff}.
\end{proof}

\section{Adjustments for Discontinuous CIFs}\label{appendix:remark discontinuous}
As mentioned in Remark \ref{rem:cifjumps}, it is possible to extend our theoretical results to CIFs that are not continuous. This extends the applicability of our approach, because tied observation times occur frequently in medical applications. Often the reason is a discretization of the timescale, due to measurements being taken on a daily or weekly basis. Therefore the assumption that the CIFs only have finitely many jumps is in most cases not restrictive for applications.

We first introduce some additional notation. Denote by $A^{(j)}_k(t), A^{(j)}(t)$ the cause-specific and all-cause cumulative hazard rates. Also $J^{(j)}=\{s\in [t_1,t_2]: \Delta A(s)^{(j)}>0\}$ denotes the set of discontinuouity points of the cause-specific cumulative hazard rates. As demonstrated by \citet{dobler2017discontinuity}, with a minor correction of \citet{dobler2024erratum}, the weak convergence of the Aalen-Johansen processes $W_{n_j}^{(j)}(t)$ and therefore also the group size adjusted processes  $V_{n_j}^{(j)}(t)$ to Gaussian processes $\Tilde{\mathbb{G}}^{(j)}$ still holds on $D([t_1,t_2]).$ However, the covariance function of $\Tilde{\mathbb{G}}^{(j)}$ is more involved and is given by
    \begin{equation*} \label{disc_covfunction}
    \begin{split} 
    \Gamma^{(j)}_{\Tilde{\mathbb{G}}}(s,t) = & \frac{1}{\kappa^{(j)}}\Bigg(\int_{0}^{s \wedge t}\frac{(S_2^{(j)}(u)-F_1^{(j)}(t))(S_2^{(j)}(u)-F_1^{(j)}(s))}{y^{(j)}(u)}\frac{dA_1^{(j)}(u)}{1-\Delta A^{(j)}(u)}
    \\
        &+ \int_{0}^{s \wedge t}\frac{(F_1^{(j)}(u)-F_1^{(j)}(t))(F_1^{(j)}(u)-F_1^{(j)}(s))}{y^{(j)}(u)}\frac{dA_2^{(j)}(u)}{1-\Delta A^{(j)}(u)}
        \\
        &+\sum_{u\in J^{(j)}, u\leq s,t}\frac{(S^{(j)}(u))^2}{y^{(j)}(u)}\frac{\Delta A_1^{(j)}(u)\Delta A_2^{(j)}(u)}{(1-\Delta A^{(j)}(u))^2}\Bigg), ~ s,t \in [t_1,t_2].
    \end{split}
    \end{equation*}
It is apparent from the structure of the covariance function that the limiting process no longer has continuous sample paths. \citet{dobler2017discontinuity} also established, that the wild bootstrap processes have to be adjusted in order to reproduce the correct covariance structure. Therefore let $G^{(j)}_{k,l,i}, j,k,l=1,2, i=1,\dots,n_j$ be rowwise-i.i.d. random variables with $E(G^{(1)}_{1,1,1})=0, \text{Var}(G^{(1)}_{1,1,1})=1$ and $E(G^{4~(1)}_{1,1,1})<\infty.$ Denote by $\Hat{A}^{(j)}_k(t)=\int_0^t\frac{1}{Y^{(j)}(s)}dN_k^{(j)}(s), \Hat{A}^{(j)}(t)=\int_0^t\frac{1}{Y^{(j)}(s)}dN^{(j)}(s)$ the cause-specific and all-cause Nelson-Aalen estimators. The adjusted wild bootstrap processes are given by
\begin{equation*}
     \label{discadjusted_bootprocess}
    \widehat{\widetilde{V}}^{(j)}_{n_j}(t)= \int_{0}^{t}\frac{\hat{S}^{(j)}_2(u)-\hat{F}^{(j)}_1(t)}{1-\Delta \Hat{A}^{(j)}(u)}d\widehat{W}^{(j)}_{1}(u) +
    \int_{0}^{t}\frac{\hat{F}^{(j)}_1(u)-\hat{F}^{(j)}_1(t)}{1-\Delta \Hat{A}^{(j)}(u)}d\widehat{W}^{(j)}_{2}(u)
\end{equation*}
with
\begin{multline*}
\widehat{W}^{(j)}_k(t) =
\sqrt{n}\sum_{i=1}^{n_j}G^{(j)}_{k,k,i}\int_{0}^{t}\sqrt{1-\Delta\widehat{A}(u)}\,\frac{dN^{(j)}_{k,i}(u)}{Y^{(j)}(u)}
+\sqrt{\frac{n}{2}}\sum_{l=1}^{k}\operatorname{sign}(l-k)
\\
\times \sum_{i=1}^{n_j}\Bigl[
G^{(j)}_{k,l,i}\int_{0}^{t}\sqrt{\Delta\widehat{A}^{(j)}_k(u)}\,\frac{dN^{(j)}_{l,i}(u)}{Y^{(j)}(u)}
+G^{(j)}_{l,k,i}\int_{0}^{t}\sqrt{\Delta\widehat{A}^{(j)}_{l}(u)}\,\frac{dN^{(j)}_{k,i}(u)}{Y^{(j)}(u)}
\Bigr]
\end{multline*}
where $\operatorname{sign}(x)=1\{x>0\}-1\{x<0\}$ is the signum function. 

As our proofs regarding (conditional) weak convergence only rely on the convergence of the Aalen-Johansen processes and their wild bootstrap counterparts, all statements remain correct for the discontinuous case, as long as the adjusted wild bootstrap processes are used. Only the proof of Theorem \ref{consistency_test_eff} made explicit use of continuity, in order to establish the consistency of our test on the {closed} interval $[t_1,t_2].$ If we exclude the right endpoint $t_2$ from our hypothesis, the proof remains valid in the discontinuous case as well. This restriction is also rather intuitive, because CIFs with $F^{(1)}_1(t_2)\ne F^{(2)}_1(t_2)$ and $F^{(1)}_1(t) = F^{(2)}_1(t)$ for all $t \in [t_1, t_2)$ lie in the same equivalence class as elements of $L^1([t_1,t_2])$ whose norm our test statistic is based upon.

\section{Additional Simulation Results}\label{appendix:add simus}

\subsection{Further Plots and Tables}\label{ssec: add plots}
\begin{table}[ht]
\centering\caption{Censoring rates for the different simulation setups}
\begin{tabular}{cclrr}
  \hline
model & hypothesis & censoring parameters & censoring rate (group 1) & censoring rate (group 2) \\ 
  \hline
1 & $H_0$ & 0;0 & 0.00 & 0.00 \\ 
  1 & $H_0$ & 0.33;1.6 & 0.25 & 0.30 \\ 
  1 & $H_0$ & 0.67;1.6 & 0.40 & 0.30 \\ 
  1 & $H_0$ & 1;1.6 & 0.50 & 0.30 \\ 
  1 & $H_0$ & 0.33;2.5 & 0.25 & 0.20 \\ 
  1 & $H_0$ & 0.67;2.5 & 0.40 & 0.20 \\ 
  1 & $H_0$ & 1;2.5 & 0.50 & 0.20 \\ \hline
  1 & $H_1$ & 0;0 & 0.00 & 0.00 \\ 
  1 & $H_1$ & 0.33;1.6 & 0.25 & 0.30 \\ 
  1 & $H_1$ & 0.67;1.6 & 0.40 & 0.30 \\ 
  1 & $H_1$ & 1;1.6 & 0.50 & 0.30 \\ 
  1 & $H_1$ & 0.33;2.5 & 0.25 & 0.20 \\ 
  1 & $H_1$ & 0.67;2.5 & 0.40 & 0.20 \\ 
  1 & $H_1$ & 1;2.5 & 0.50 & 0.20 \\ \hline
  2 & $H_0$ & 0;0 & 0.00 & 0.00 \\ 
  2 & $H_0$ & 0.33;1.6 & 0.29 & 0.66 \\ 
  2 & $H_0$ & 0.67;1.6 & 0.49 & 0.66 \\ 
  2 & $H_0$ & 1;1.6 & 0.63 & 0.66 \\ 
  2 & $H_0$ & 0.33;2.5 & 0.29 & 0.43 \\ 
  2 & $H_0$ & 0.67;2.5 & 0.49 & 0.43 \\ 
  2 & $H_0$ & 1;2.5 & 0.63 & 0.43 \\ \hline
  2 & $H_1$ & 0;0 & 0.00 & 0.00 \\ 
  2 & $H_1$ & 0.33;1.6 & 0.30 & 0.68 \\ 
  2 & $H_1$ & 0.67;1.6 & 0.50 & 0.68 \\ 
  2 & $H_1$ & 1;1.6 & 0.64 & 0.68 \\ 
  2 & $H_1$ & 0.33;2.5 & 0.30 & 0.51 \\ 
  2 & $H_1$ & 0.67;2.5 & 0.50 & 0.51 \\ 
  2 & $H_1$ & 1;2.5 & 0.64 & 0.51 \\ 
   \hline
\end{tabular}\label{tab:censrates}
\end{table}

\begin{figure}
    \centering
    \includegraphics[width=\linewidth]{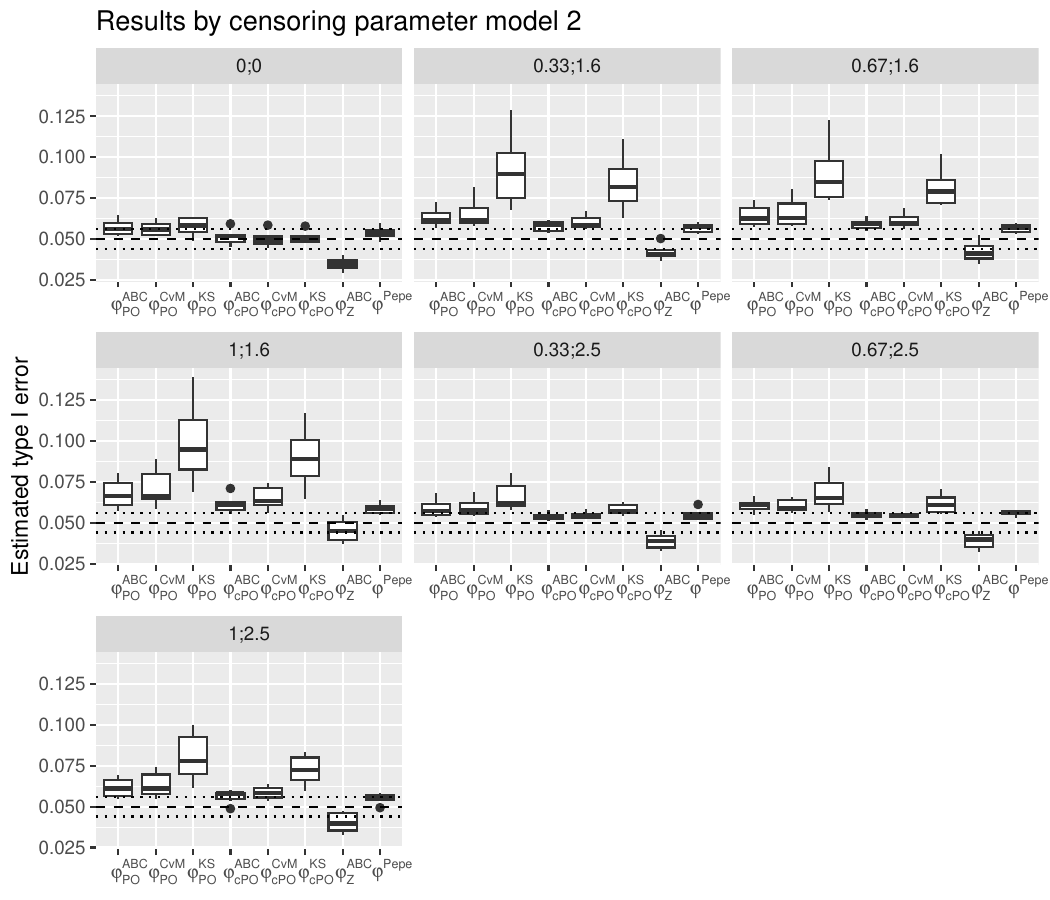}
    \caption{Estimated type I error under model 2 stratified by censoring parameters}
    \label{fig:bycens_mod2_h0}
\end{figure}

\begin{figure}
    \centering
    \includegraphics[width=\linewidth]{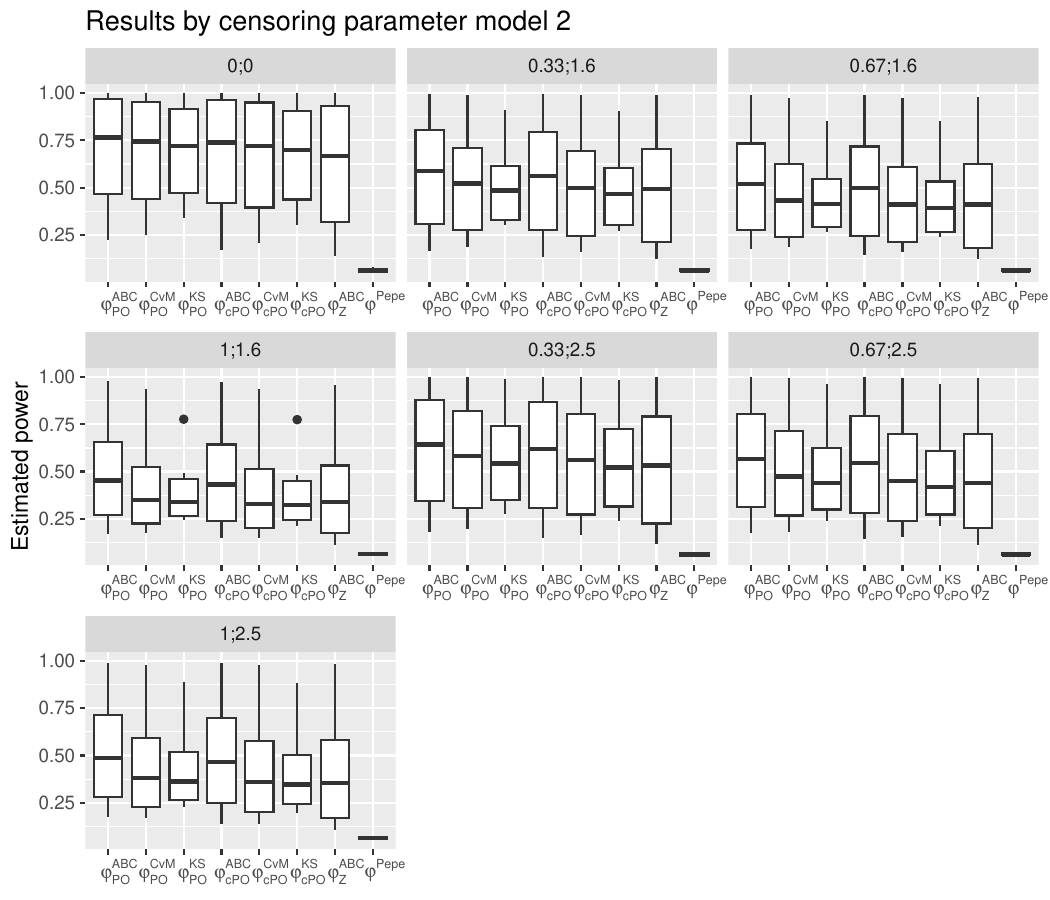}
    \caption{Estimated power under model 2 stratified by censoring parameters}
    \label{fig:bycens_mod2_h1}
\end{figure}

\FloatBarrier
\subsection{Detailed Results of Model~1}\label{ssec:model1}
\begin{figure}[h]
    \centering
    \includegraphics[width=\linewidth]{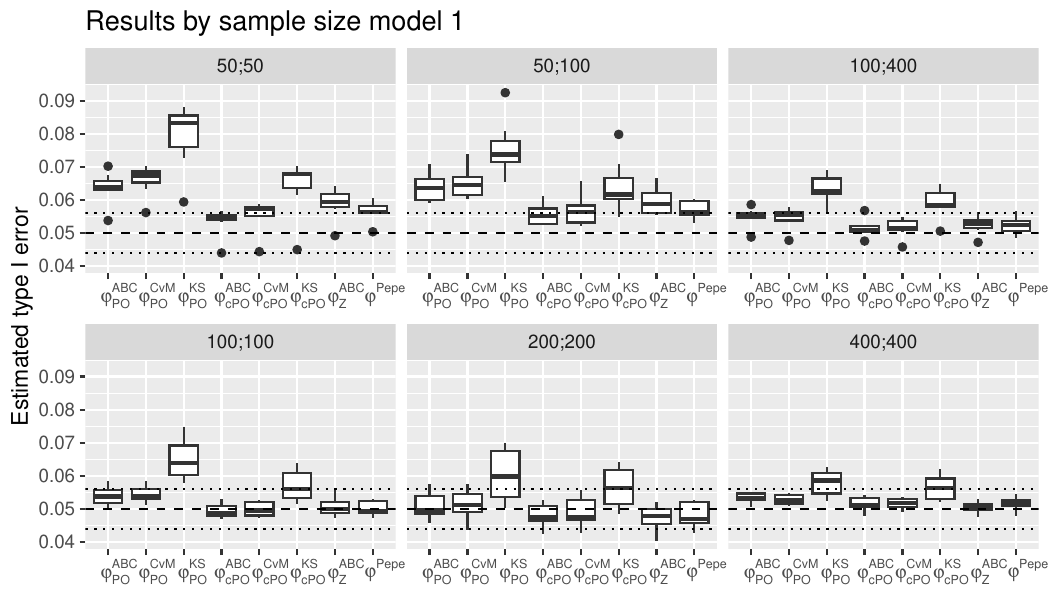}
    \caption{Estimated type I error under model 1 stratified by sample sizes}
    \label{fig:bysample_mod1_h0}
\end{figure}
\begin{figure}
    \centering
    \includegraphics[width=\linewidth]{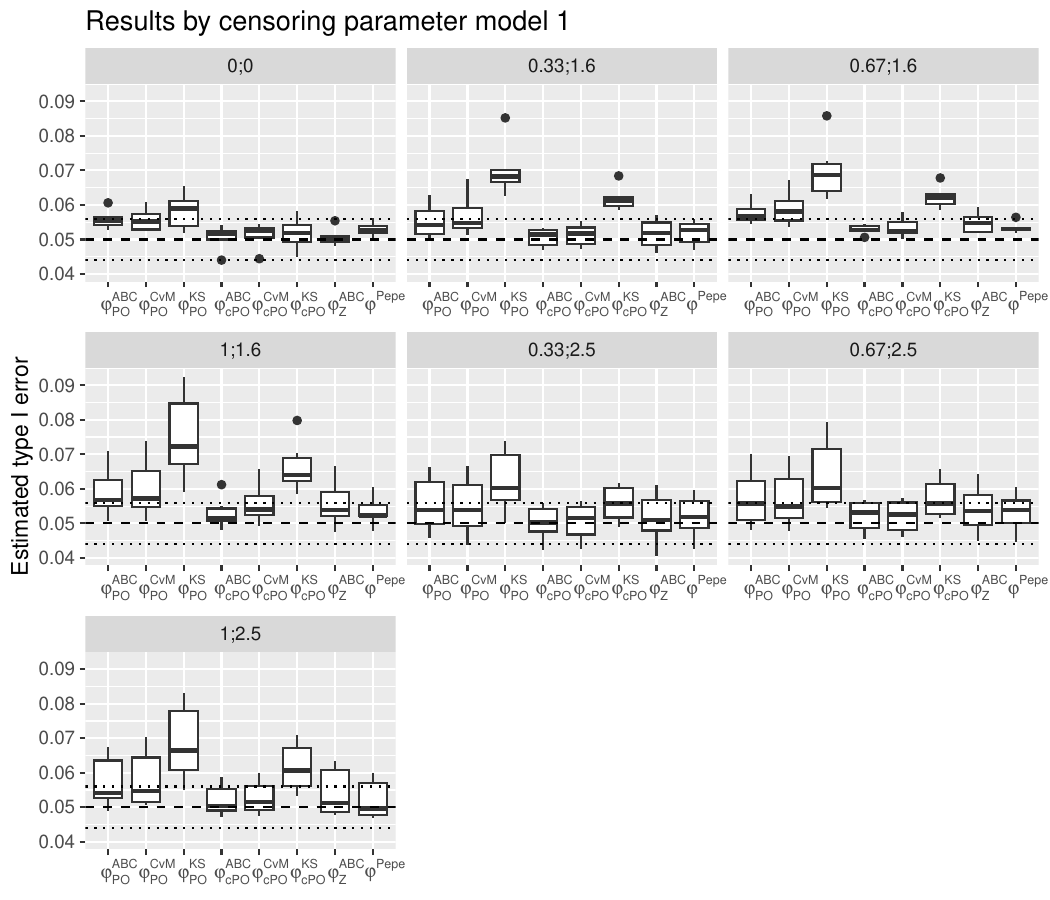}
    \caption{Estimated type I error under model 1 stratified by censoring parameters}
    \label{fig:bycens_mod1_h0}
\end{figure}

\begin{figure}
    \centering
    \includegraphics[width=\linewidth]{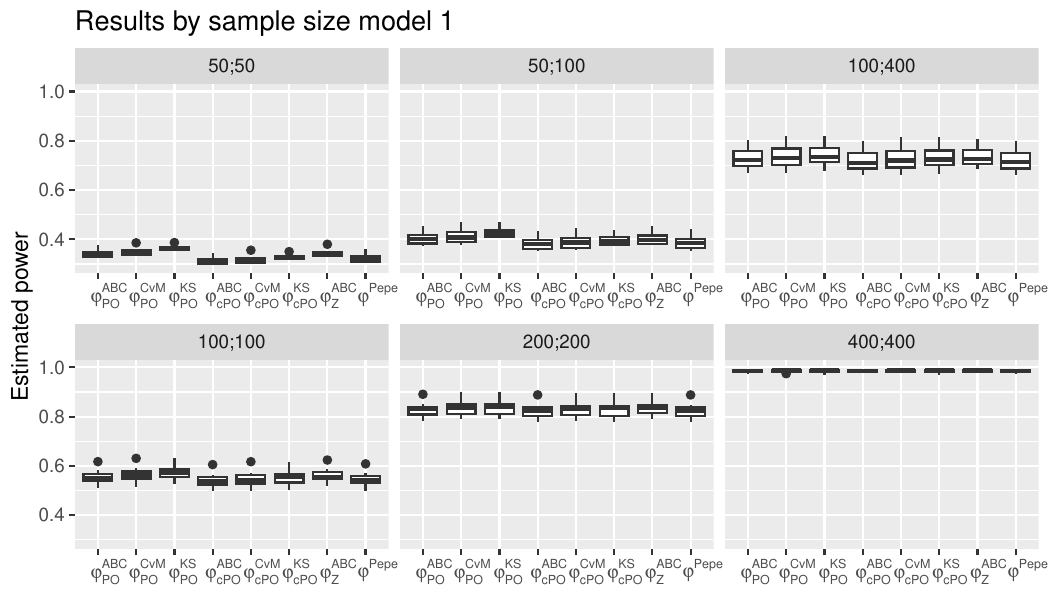}
    \caption{Estimated power under model 1 stratified by sample sizes}
    \label{fig:bysample_mod1_h1}
\end{figure}
\begin{figure}
    \centering
    \includegraphics[width=\linewidth]{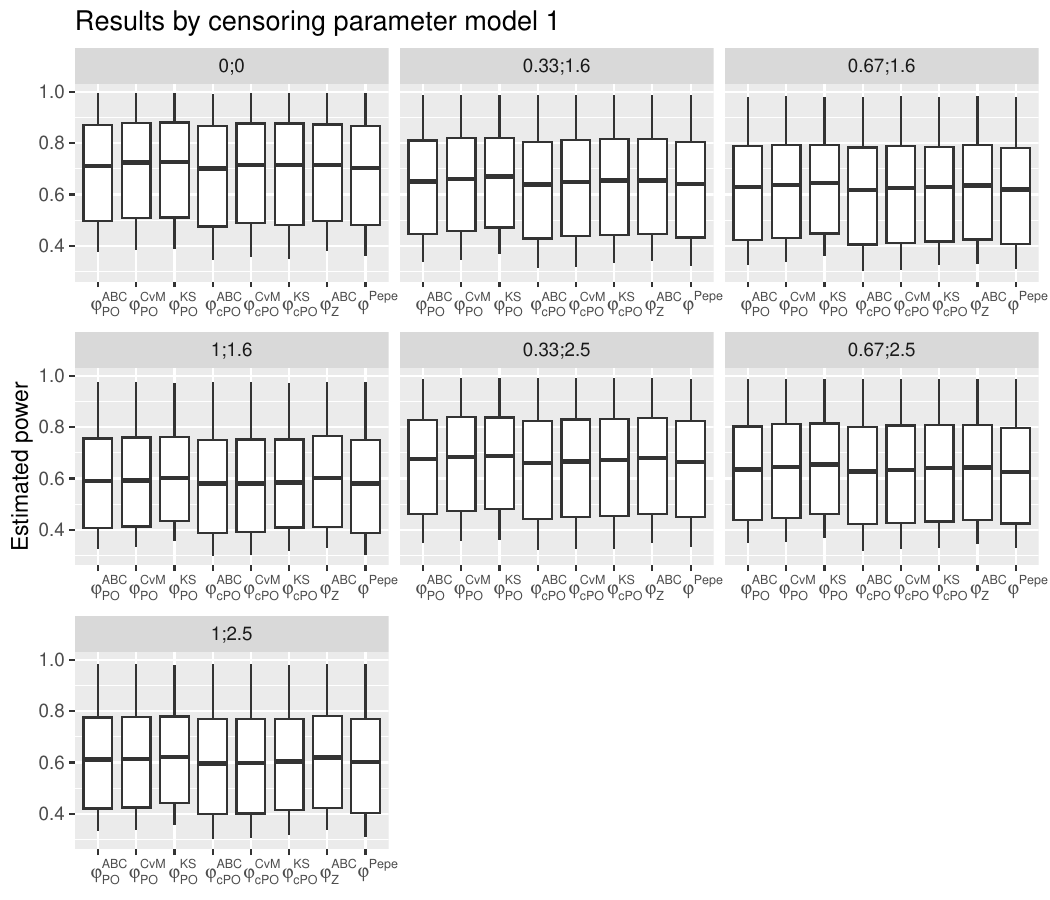}
    \caption{Estimated power under model 1 stratified by censoring parameters}
    \label{fig:bycens_mod1_h1}
\end{figure}
\FloatBarrier

\bibliographystyle{abbrvnat}
\bibliography{literatur}

\end{document}